\documentstyle[aps,epsfig]{revtex}

\begin{document}


\title{Storage capacity of a constructive learning algorithm}
\author{Arnaud Buhot and Mirta B. Gordon~\footnote{ also 
at Centre National de la Recherche Scientifique}}
\address{SPSMS/DRFMC/CEA Grenoble, 17 av. des Martyrs, 
38054 Grenoble Cedex 9, France}

\maketitle

\begin{abstract}
Upper and lower bounds for the typical storage capacity of a 
constructive algorithm, the Tilinglike Learning Algorithm for the 
Parity Machine [M. Biehl and M. Opper, Phys. Rev. A {\bf 44} 
6888 (1991)], are determined in the asymptotic limit 
of large training set sizes. The properties of a 
perceptron with threshold, learning a training set of patterns 
having a biased distribution of targets, needed 
as an intermediate step in the capacity calculation, are 
determined analytically. The lower bound for the capacity, 
determined with a cavity method, is proportional to the 
number of hidden units. The upper bound, obtained with 
the hypothesis of replica symmetry, is close to the 
one predicted by Mitchinson and Durbin [Biol. 
Cyber. {\bf 60} 345 (1989)].    
\end{abstract}

\section{Introduction}

In this paper, we consider the problem of 
learning binary classification tasks from 
examples with neural networks. The network's 
architecture and the neurons' weights are 
determined based on a training set of examples 
or patterns ${\cal L}_{\alpha}$, composed of 
$P= \alpha N$ input vectors $\left\{{\bf x}^{\mu} 
\right\}_{\mu=1,\cdots,P}$ in $N$-dimensional space
and their corresponding 
classes $\tau^{\mu} = \pm 1$. The latter are the targets to 
be learned. Hereafter, we call $\alpha \equiv P/N$ the 
{\it size} of the training set.  
One interesting property that characterizes a neural 
network is its storage capacity, which is the 
size $\alpha_c$ of the largest training set with 
arbitrary targets the network is able to 
learn (with probability $1$). 
The perceptron, a single neuron connected to its 
inputs through $N$ weights, performs linear separations 
and has a storage capacity $\alpha_c = 
2$~\cite{Cover,Gardner,Gardner2,GaDe}. 
It is possible to increase the storage capacity 
of neural networks by considering more complicated  
architectures, like those with one hidden layer of 
$k$ units. Such monolayer perceptrons map each input 
vector ${\bf x}$ to a binary $k$-dimensional internal 
representation determined by the outputs of $k$ 
perceptrons, which in this context are also called 
hidden units. The overall network's output to an input 
pattern is a boolean function of the corresponding 
internal representation. 
This function may be learned by an output perceptron, 
but then the internal representations of the 
training set must be linearly separable. 
In order to get rid of this constraint, networks 
implementing particular functions of the hidden 
states have been investigated. Among these, the 
committee machine, whose output is the class of 
the majority of the hidden units, and the parity 
machine, whose output is the 
product of the $k$ components of the internal 
representation, have deserved particular 
attention~\cite{WRB}. 

Learning consists of adapting the number of 
hidden perceptrons and their weights in order 
that the outputs of the network to the 
training examples match the corresponding targets. 
The main problem is that the internal representations 
are unknown. Besides the CHIR algorithm~\cite{GMD}, 
that determines the internal representations through a 
random process involving learning faithful sets 
of internal representations with $k$ 
fixed, most learning algorithms build the internal 
representations through a deterministic incremental 
procedure that determines $k$ by construction. In the 
latter case, the hidden perceptrons are trained 
one after the other with targets that differ 
from one algorithm to another,
until the correct classification is achieved. 
The first incremental procedure 
has been proposed by Gallant~\cite{Gallant}. 
Many other authors developed further this idea, like 
M\'ezard and Nadal with the Tiling Algorithm~\cite{MeNa}, 
Ruj\'an and Marchand with the Sequential Learning 
Algorithm~\cite{RuMa} and Biehl and Opper with the 
Tilinglike Learning Algorithm~\cite{BiOp}.  
Other variations have been proposed~\cite{Frean,MaEs}.  
It has been argued that these incremental procedures 
may require a number of hidden units much larger 
than the number actually needed by a network making 
use of its full storage capacity. In the following we 
distinguish thus the algorithm's capacity, defined as the  
size of the largest training set (with arbitrary targets) 
learnable with the algorithm, from the capacity of the network 
with the same architecture. 
Clearly the former cannot be larger than the latter. 
An upper bound 
for the storage capacity of the parity
machine with $k$ hidden perceptrons has been obtained 
by Mitchinson and Durbin~\cite{MiDu} through a geometric 
approach: $\alpha_c(k) \leq k \ln k /\ln2$.
Recent replica calculation results, obtained 
in the limit of a large number of hidden perceptrons ($k 
\rightarrow + \infty$)~\cite{XKO}, strongly 
suggest that this upper bound may effectively be 
reached. However, the 
learning problem remains: is there a learning algorithm 
whose capacity saturates this bound?
This question was addressed recently in~\cite{WeSa2} 
within the same statistical mechanics framework as 
the present work. In spite of a thorough analysis, 
no clear-cut conclusion could be drawn in the 
asymptotic regime of large $k$, because of a lack 
of precision in the numerical integration of the 
corresponding equations. 

In this paper, we determine analytically the storage
capacity of a parity machine built with the Tilinglike
Learning Algorithm (TLA).
Our results present strong evidence showing that the
storage capacity of the obtained network is close to
the upper bound, at least within the replica-symmetry
approximation. The paper is organized as follows:
in section \ref{sec:TLA}, we describe the TLA. The 
conditions necessary for the TLA to converge impose strong 
constraints on the cost function used to train the 
hidden perceptrons. These are discussed in section 
\ref{sec:convcond}. Despite intensive 
research in this field, no analytic results on 
the learning properties of the perceptron with 
threshold, in the asymptotic limit $\alpha \rightarrow + \infty$
needed here, exist. These are deduced in section 
\ref{sec:percept} for the Gardner cost function with 
vanishing and finite margin, within
the replica-symmetry (RS) approximation.
As this approximation is known to provide only a
lower bound to the perceptron's actual training 
error~\cite{ErTh,MEZ},
we also determined an upper bound through a generalization
of the Kuhn-Tucker (KT) cavity method proposed by Gerl and
Krey~\cite{GeKr}. The general expression for 
the number of hidden perceptrons generated by 
the TLA in the limit $\alpha \rightarrow + \infty$ 
is deduced in section \ref{sec:k(alpha)}. Our main 
result is that the number of hidden units 
needed by the TLA to converge grows proportionally to 
$\alpha/(\ln \alpha)^\nu$ in the large $\alpha$ limit, 
where $\nu =1$ in the RS approximation and $\nu=0$ within 
the KT cavity method, provided that the hidden perceptrons
learn through the minimization of 
their training errors.
Our results are discussed and compared both to the 
Mitchinson and Durbin bound~\cite{MiDu} and to 
the numerical results obtained by West and 
Saad~\cite{WeSa2}. The general conclusion is left 
to section \ref{sec:concl}.

\section{The Tilinglike Learning Algorithm (TLA)}
\label{sec:TLA}

In the following, we describe the Tilinglike 
Learning Algorithm (TLA) considered in the following 
because of its simplicity. The TLA needs hidden 
perceptrons with a threshold to generate the parity 
machine. The classification performed by a perceptron 
is a linear separation defined by a hyperplane in 
the $N$-dimensional input space, of 
normal vector ${\bf J}$ (${\bf J} \cdot {\bf J} = 1$) 
and distance to the origin $\theta$. The $N$ components of ${\bf J}$ are the 
perceptron's weights and $\theta$ is its threshold. 
An example ${\bf x}$ is classified as follows:
 
\begin{equation}
\label{classe}
\sigma \equiv {\rm sign} \left( {\bf J} 
\cdot {\bf x} - \theta \right).
\end{equation}
 
As already pointed out in~\cite{WeSa2} the threshold 
is useful in the case of unbalanced training sets, 
containing more examples of one class than of the other. 
As we will see in the following, this is the case for 
the successive perceptrons included by the TLA.
 
In the first learning step of the algorithm, the 
parameters ${\bf J}_1$ and $\theta_1$ of a perceptron 
are adapted in order to obtain the lowest possible 
number of training errors. 
This is usually done through the minimization 
of a cost function:

\begin{equation}
\label{cost}
E( {\bf J}_1,\theta_1;{\cal L}_{\alpha}) =   
\sum_{\mu = 1}^{P} V \left( \lambda_1^{\mu} \right)
\end{equation}

\noindent where the potential $V$ is a function of 
$\lambda_1^{\mu}$, the stability of the example $\mu$:

\begin{equation}
\label{stab}
\lambda_1^{\mu} \equiv \tau^{\mu} \left( {\bf J}_1 
\cdot {\bf x}^{\mu} - \theta_1 \right).
\end{equation}

\noindent The stability is positive if and
only if the example is correctly classified.
Its absolute value is the distance of the 
example to the separating hyperplane.

In principle, there is some freedom in the choice 
of the potential $V(\lambda)$. 
As it has to penalize training errors, it has to be 
a decreasing function of $\lambda$. 
Considering as cost function the number of 
training errors corresponds to the particular choice $V(\lambda) 
= \Theta(-\lambda)$, where $\Theta(x)$ is the Heaviside 
function. Other potentials, that do not minimize the 
number of training errors but possess interesting 
learning or algorithmic properties may be chosen. 
Examples are $V(\lambda) = (\kappa - \lambda)^{\, n} \,  
\Theta (\kappa - \lambda)$ where $\kappa \geq 0$ 
is a fixed positive margin chosen {\it a priori}.
The case $n=0$ corresponds to the so-called Gardner 
potential~\cite{Gardner2,GaDe} which reduces to the 
error counting function for $\kappa =  0$. 
The potential defined by $n=1$ corresponds to the 
Perceptron learning algorithm~\cite{AbKe,AbKe2,GrGu} 
and $n=2$ to the AdaTron~\cite{AnBi,GrGu}. 

After learning, the training error of the first 
perceptron is:

\begin{equation}
\varepsilon_t^1 ({\bf J}_1^*,\theta_1^*;{\cal L}_{\alpha}) 
= \frac{1}{P} \sum_{\mu = 1}^{P} \Theta \left(- \tau^{\mu} 
\sigma_1^{\mu} \right)
\end{equation}

\noindent where $\sigma_1^{\mu}$, the class given by 
the perceptron to the example ${\bf x}^{\mu}$, depends 
through equation~(\ref{classe}) on the parameters ${\bf J}_1^*$ 
and $\theta_1^*$ that minimize the cost function.

If the training error is zero, the learning 
procedure stops. Then, the class associated 
to the patterns by the parity machine is just 
the class given by the first perceptron. 
Otherwise, another perceptron is included and 
trained with the aim of separating the correctly 
learned examples from the wrongly learned ones. 
The corresponding training set ${\cal L}_{\alpha}^2 
= \left\{ {\bf x}^{\mu}, \tau_2^{\mu} \right\}_{\mu 
= 1, \cdots,P}$ contains the same input examples as 
${\cal L}_{\alpha}$ with new targets $\tau_2^{\mu}$ 
defined as follows: $\tau_2^{\mu} = + 1$ if the example 
${\bf x}^{\mu}$ is correctly classified by the previous 
perceptron and $\tau_2^{\mu} = - 1$ if not.
These targets may be expressed as $\tau_2^{\mu} = 
\sigma_1^{\mu} \tau^{\mu}$. 
Notice that a fraction $1 - \varepsilon_t^1$ of patterns 
have targets $+1$, and a fraction $\varepsilon_t^1$ 
have targets $-1$. Since we expect the training error 
$\varepsilon_t^1$ to be smaller than $1/2$, the probability 
of targets $-1$ is smaller than that of targets $+1$. 
The successive perceptrons need a threshold to 
learn such biased training sets. Otherwise, the tilinglike 
construction cannot converge. 

The parameters ${\bf J}_2^*$ and $\theta_2^*$ 
of the second perceptron are learned with the 
training set ${\cal L}_{\alpha}^2$, minimizing 
the same cost function as the first one. 
The same procedure, in which the perceptron $i+1$ 
learns the training set ${\cal L}_{\alpha}^{i+1} = 
\left\{ {\bf x}^{\mu},\tau_{i+1}^{\mu} = \tau_i^{\mu} 
\sigma_i^{\mu} \right\}_{\mu = 1,\cdots,P}$,  
has to be iterated until $\varepsilon_t^k = 0$. 
Then, the product $\sigma^{\mu}$ of the classes 
$\left\{ \sigma_i^{\mu} \right\}_{i=1,\cdots,k}$ 
given by the hidden perceptrons to an example 
${\bf x}^{\mu}$ corresponds to the target 
$\tau^{\mu}$~\cite{BiOp,MaEs}, as $\sigma^{\mu} \equiv 
\sigma_1^{\mu} \cdots \sigma_k^{\mu} = \sigma_1^{\mu} 
\cdots \sigma_{k-2}^{\mu} (\sigma_{k-1}^{\mu})^2 
\tau_{k-1}^{\mu} = \sigma_1^{\mu} \cdots \sigma_{k -
2}^{\mu} \tau_{k-1}^{\mu} = \cdots = \tau^{\mu}$. 
Thus, the TLA constructs a parity machine with $k$ 
hidden units.

\section{Convergence conditions}
\label{sec:convcond}

It has been shown that if the examples are 
binary~\cite{BiOp}, or real-valued vectors in 
general position~\cite{Gordon}, there is a 
solution that satisfies the TLA construction 
with the property that $P \varepsilon_t^i$ 
is a succession  of decreasing integer numbers. 
Thus, a finite $k \leq P$ exists for which 
$\varepsilon_t^k = 0$.

In the following, we are interested in the {\it typical} number 
$k$ of hidden perceptrons necessary for the TLA to learn 
a training set of size $\alpha$. This is obtained in the 
thermodynamic limit where $N$ and $P$ diverge 
keeping $\alpha = P/N$ constant. 
In this limit, $k$ is expected to be independent 
of the particular set of training patterns, and 
to depend only on $\alpha$. 
However, as $P \rightarrow + \infty$, it is not possible 
to argue that $P \varepsilon_t^i$ is a succession 
of strictly decreasing numbers in order to guarantee 
the convergence of the TLA in a finite number of 
steps (i.e. of hidden units).
In particular, the solution in which a single 
example is correctly learned at each step, used
by the convergence proofs~\cite{BiOp,Gordon} at
finite $N$, leads to $k \rightarrow +\infty$. 
In order to obtain a {\it finite} number $k(\alpha)$ in 
the thermodynamic limit, each perceptron has to learn 
at least a number of examples of the order of $N$. 
This imposes some general conditions on the 
learning algorithm used to train the perceptrons.

It is worth to point out that the conditions for 
convergence with finite $k$ in the thermodynamic 
limit do not guarantee the convergence for 
{\it all} the possible training sets of size $\alpha$.
This is due to the probabilistic nature of the
statistical physics results, which predict the average
behaviour. The results may not be correct for subsets of zero
measure in the space of training sets, and in particular 
for the worst case.
 
As described before, the training set 
${\cal L}_{\alpha}^i$ used to train the perceptron $i$ 
contains a fraction $1-\varepsilon_t^{i-1}$ 
of patterns with targets $+1$, and a fraction 
$\varepsilon_t^{i-1}$ of patterns with targets $-1$. 
These targets are slightly correlated, as 
they are determined by the training errors 
of the preceding perceptron. However, it has been 
shown that these correlations are weak~\cite{WeSa2}. 
We neglect them in the limit $\alpha \rightarrow 
+ \infty$ considered in the following. 
Thus, we consider that the targets to be 
learned by the successive perceptrons are i.i.d. 
random variables, and have a probability 
$1 - \varepsilon_t^{i-1}$ to be $+1$ and
$\varepsilon_t^{i-1}$ to be $-1$. 
As this neglects the constraints 
imposed by the correlations on the minimization 
of the training error, we expect that the  
assumption of uncorrelated targets underestimate the 
perceptrons' training errors. 
It follows that our estimation of the number 
$k(\alpha)$ of perceptrons necessary to construct 
the parity machine is a lower bound to the actual value.

Consider a perceptron learning a training 
set of size $\alpha$ with targets 
given by the following biased probability distribution: 

\begin{equation}
\label{prob}
P(\tau) = (1-\varepsilon) \, \delta(\tau-1) + 
\varepsilon \, \delta (\tau +1). 
\end{equation}
 
\noindent If ${\cal E}_t (\alpha, \varepsilon)$ 
is the perceptron's training error, i.e. the fraction of 
wrongly learned examples, there is a simple relationship between 
the training errors $\varepsilon_t^{i-1}$ and 
$\varepsilon_t^i$ of two successive hidden perceptrons: 

\begin{equation} 
\label{relationEt} 
\varepsilon_t^i = {\cal E}_t (\alpha, \varepsilon_t^{i-1})
\end{equation}
  
\noindent since the bias in the probability of the 
targets of perceptron $i$ is due to the training 
error of the preceding unit. 

The successive training errors 
$\varepsilon_t^i$ must decrease monotonically 
with $i$ and eventually vanish for a finite $k$. 
Otherwise the TLA does not converge.
Taking equation~(\ref{relationEt}) into account, 
this imposes that: 

\begin{equation}
\label{condition}
{\cal E}_t (\alpha, \varepsilon) < \varepsilon.
\end{equation}

\noindent Condition~(\ref{condition}) 
restricts the possible potentials in the cost 
function~(\ref{cost}). For example, in the 
following section we show that the Perceptron and 
the AdaTron potentials~\cite{AbKe,AbKe2,GrGu,AnBi} 
do not satisfy the condition~(\ref{condition}) 
for all $\alpha$ when $\varepsilon < 1/2$. 

The stopping condition of the TLA  
imposes that there is a finite value of $k$ such that:

\begin{equation}
\label{stop}
\varepsilon_t^k = {\cal E}_t (\alpha,\varepsilon_t^{k-1}) = 0.
\end{equation}

\noindent This in turn imposes that for all $\alpha$, 
there always exists $\varepsilon_0 (\alpha) \neq 0$  
such that ${\cal E}_t (\alpha,\varepsilon_0 
(\alpha)) = 0$. Thus, the stopping condition~(\ref{stop}) 
imposes that the inverse function $\alpha_0(\varepsilon)$ 
diverges as $\varepsilon \rightarrow 0$.  
In fact, $\alpha_0(\varepsilon)$ is  the storage capacity 
of a perceptron learning targets drawn with the biased 
probability~(\ref{prob}) (in the literature, the bias is 
usually defined as $1 - 2 \varepsilon$).  
Actually, the divergence of $\alpha_0(\varepsilon)$ 
occurs whenever the potential $V(\lambda)$ vanishes for 
$\lambda > 0$ and is strictly positive for $\lambda < 0$.
This is the case for the Gardner potential with $\kappa 
= 0$, for which $\alpha_0 (\varepsilon) \sim - (\varepsilon 
\ln \varepsilon)^{-1}$~\cite{Gardner,Gardner2,GaDe}.
However, even if the perceptron has been extensively studied, 
very few results exist for the case of training sets with 
biased distributions of targets~\cite{Gardner2,BiOp,WeSa}. 
In particular, the asymptotic behaviour of the learning curves 
${\cal E}_t (\alpha, \varepsilon)$ as a function of $\alpha$ 
is unknown. These are deduced in the next section. 
The reader not interested in these intermediate 
calculations may skip them and go straight 
to section \ref{sec:k(alpha)}. Only the results displayed by 
equations~(\ref{EqEt}), (\ref{EqEtkappa}), 
(\ref{eqEtKT}) and (\ref{eqEtKTkappa}) are used 
to determine the asymptotic behaviour of the TLA. 

\section{Perceptron's training error for biased target-distributions}
\label{sec:percept}

In order to learn such training sets with 
biased distributions of targets, the perceptron must have 
a threshold, as the separating hyperplanes that 
minimize the training error do not contain the origin. 
Here we present new analytic 
results, mainly in the asymptotic regime $\alpha 
\rightarrow +\infty$, for the Gardner 
cost function defined by the potential: 

\begin{equation}
\label{potential}
V (\lambda) = \Theta ( \kappa - \lambda). 
\end{equation}

\noindent For $\kappa = 0$, the corresponding cost 
function is the number of training errors. For $\kappa 
> 0$, the cost function is the number of examples 
with stability~(\ref{stab}) smaller than $\kappa$.

The section is divided in two parts. In the first 
one we derive results within the Replica-Symmetry 
(RS) approximation, which is known to underestimate 
the training error. In the second part we obtain upper 
bounds for the training error, using a cavity method.

\subsection{Replica calculation}

We briefly recall the main steps of the replica calculation, 
that follows the same lines as~\cite{BiOp,WeSa}. 
As we are interested in the properties of the minimum of 
the cost function, a temperature $T \equiv 1/\beta$ is 
introduced and the cost function is considered as 
an energy. The corresponding partition function writes:

\begin{equation} 
\label{partition}
Z(\beta, {\cal L}_{\alpha}(\varepsilon)) = \int 
d \theta P(\theta) \, \int d {\bf J} 
P({\bf J}) \exp \left( - \beta 
E({\bf J},\theta;{\cal L}_{\alpha}(\varepsilon) \right)
\end{equation} 
 
\noindent where the components of ${\bf J}$ 
are the weights, and $\theta$ is the perceptron's 
threshold. ${\cal L}_{\alpha} (\varepsilon)$ is a training set 
of size $\alpha$. The input vectors ${\bf x}^{\mu}$ 
are drawn from a gaussian distribution with zero mean and unit 
variance in all the directions. The targets have the biased
distribution (\ref{prob}). 

Following Gardner's approach, the patterns of the training 
set are considered as frozen disordered variables. 
The replica trick allows to calculate the mean free 
energy in the thermodynamic limit ($N \rightarrow + \infty$, 
$P \rightarrow + \infty$ and $\alpha$ constant) averaged 
over all possible training sets, as follows: 

\begin{equation} 
f(\alpha,\varepsilon) = \lim_{\beta \rightarrow + \infty}
\lim_{{\scriptstyle N \rightarrow +\infty \atop \scriptstyle
P \rightarrow +\infty} \atop \scriptstyle \alpha = P/N}
\lim_{n \rightarrow 0} - \frac{1}{\beta n N} \ln \overline{Z^n
(\beta, {\cal L}_{\alpha}(\varepsilon))}
\end{equation}

\noindent where the bar stands for the mean over  
the training sets with same size $\alpha$. Thus, 
the free energy is obtained through the averaging 
of a partition function of $n$ replicas of the 
original system. Hereafter we assume replica symmetry 
(RS), i.e. that the replicas are equivalent 
under permutation. However, it is well known that replica 
symmetry breaks down when the training error is 
finite~\cite{Bouten}. Calculations including one 
step of replica symmetry breaking have shown that the
training error obtained within the RS approximation is
a lower bound for the actual one~\cite{ErTh,MEZ}. 
 
Assuming that the weights have a uniform  prior 
probability over the surface of the $N$-dimensional 
sphere of unitary radius, and the threshold a uniform 
distribution over the real axis between $-\sqrt{N}$ 
and $+\sqrt{N}$, the free energy within the RS 
approximation writes:

\begin{equation}
f(\alpha,\varepsilon) = \max_c \, \min_{\theta} 
\, g(\alpha,\varepsilon,c,\theta)
\end{equation}

\noindent where the function $g$ is:

\begin{eqnarray}
\nonumber g(\alpha,\varepsilon,c,\theta) & = & 
- \frac{1}{2c} + \alpha ( 1 - \varepsilon ) \int 
W(\lambda(y,c),y,c) \exp \left( - \frac{(y + 
\theta)^2}{2} \right) \frac{dy}{\sqrt{2\pi}}\\
\label{foncg}
& & + \, \alpha \, \varepsilon \int W(\lambda(y,c),y,c)
\exp \left( - \frac{(y - \theta)^2}{2} \right) 
\frac{dy}{\sqrt{2\pi}}
\end{eqnarray}

\noindent with $\lambda(y,c)$ the function that minimizes:
$W(\lambda,y,c) \equiv V(\lambda) + (\lambda - y)^2/2c$.
$c$ is the usual order parameter in replica calculations
($c = \lim_{\beta \rightarrow +\infty} \beta (1-{\bf J}_a 
\cdot {\bf J}_b)$ with ${\bf J}_a$ and ${\bf J}_b$ the 
directions corresponding to two different replicas). 
The parameters $c$ and $\theta$ are solutions 
of the following extremum conditions:

\begin{equation}
\label{foncextr}
\frac{\partial g}{\partial c} = 
\frac{\partial g}{\partial \theta} = 0.
\end{equation}

The training error ${\cal E}_t (\alpha,\varepsilon)$ 
may be easily deduced by integration of the distribution 
of stabilities over the negative values~\cite{WeSa}, 
yielding: 
 
\begin{eqnarray}
\nonumber {\cal E}_t(\alpha,\varepsilon) & = & 
\left(1 - \varepsilon \right) \int \Theta \left( - 
\lambda(y,c) \right) \exp \left( - \frac{(y + 
\theta)^2}{2} \right) \frac{dy}{\sqrt{2\pi}}\\
\label{foncEt}
& & + \, \varepsilon \int \Theta \left( - 
\lambda (y,c) \right) \exp \left( - \frac{(y - 
\theta)^2}{2} \right) \frac{dy}{\sqrt{2\pi}}.
\end{eqnarray}

Equations (\ref{partition}) to (\ref{foncEt}) are
valid for any potential $V(\lambda)$ in (\ref{cost}).
In the following, we concentrate specifically on the 
Gardner potential~(\ref{potential}). 
The function $\lambda (y,c)$ that minimizes  
$W(\lambda,y,c)$ for a given $\kappa$ is:

\begin{equation}
\label{lambda}
\lambda(y,c)  = \left\{ 
\begin{array}{cccrcl}
y & & {\rm for} & &y&< \kappa - \sqrt{2c}\\
\kappa & & {\rm for} & \kappa - \sqrt{2c} <&y&< \kappa\\
y & & {\rm for} & \kappa <&y& \\
\end{array}
\right.
\end{equation}

\noindent Introducing (\ref{lambda}) into (\ref{foncg}), 
we deduce $g(\alpha,\varepsilon,c,\theta)$.
The conditions (\ref{foncextr}) allow to determine the
equations for $c$ and $\theta$:

\begin{eqnarray}
\label{Eqc1}
\frac{1}{\alpha} & = & (1-\varepsilon) \int_{\kappa - 
\sqrt{2c} + \theta}^{\kappa + \theta} (\kappa + 
\theta - y)^2 Dy + \varepsilon \int_{\kappa - \sqrt{2c} 
- \theta}^{\kappa - \theta} (\kappa - \theta - y)^2 Dy,\\
\label{Eqtheta1}
0 & = & (1-\varepsilon) \int_{\kappa - \sqrt{2c} + 
\theta}^{\kappa + \theta} (\kappa + \theta - y) Dy - 
\varepsilon \int_{\kappa - \sqrt{2c} 
- \theta}^{\kappa - \theta} (\kappa - \theta - y) Dy,
\end{eqnarray}

\noindent where $Dy = \exp(-y^2/2) \, dy/\sqrt{2\pi}$.
The distribution of stabilities of the training patterns 
is $\rho(\lambda) = (1-\varepsilon) \rho_{+} (\lambda) + 
\varepsilon \rho_{-} (\lambda)$ with: 

\begin{eqnarray}
\rho_{\pm} (\lambda) & = & \delta ( \lambda - \kappa ) 
\int_{\kappa - \sqrt{2c} \pm \theta}^{\kappa \pm 
\theta} Dy \\ 
\nonumber & & + \left\{ \Theta (  \kappa - \sqrt{2c} - \lambda) 
+ \Theta (\lambda - \kappa ) \right\} \exp \left( - 
\frac{(\lambda \pm \theta)^2}{2} \right) \frac{1}{\sqrt{2 \pi}}. 
\end{eqnarray}
 
\noindent $\rho (\lambda)$ presents a two band 
structure with a gap between $\lambda_{-} = 
\kappa - \sqrt{2 c}$ and $\lambda_{+} = \kappa$. 
Notice that only if $\lambda_{-} < 0$ the lower 
band corresponds to wrongly classified patterns. 
If $\kappa > 0$, then $\lambda_{-}$ may become 
positive for sufficiently small values of $c$. 
In that case, the training error is only a 
fraction of the patterns lying in the lower band. 
Taking this into account, the training error 
${\cal E}_t (\alpha, \varepsilon)$ is~:
 
\begin{equation}
\label{Et1}
{\cal E}_t (\alpha,\varepsilon) = 
(1-\varepsilon) \int_{\max [- \theta, \sqrt{2c} - \kappa - 
\theta]}^{+ \infty} Dy + \varepsilon \int_{\max [\theta, 
\sqrt{2c} - \kappa + \theta]}^{+\infty} Dy,  
\end{equation}

We derive separately the asymptotic properties for 
$\kappa = 0$ and for $\kappa \neq 0$, for reasons 
that will become clear in the following. 

We consider first the case $\kappa = 0$. The band 
of positive stabilities starts at $\lambda_{+} 
= 0$ so that the gap, of width $\sqrt{2 c}$, lies 
strictly in the region of negative 
stabilities. As we expect that the gap vanishes for 
$\alpha \rightarrow + \infty$, we look for solutions 
of the extremum equations with $c \rightarrow 0$ 
and $| \theta | \rightarrow + \infty$ 
(notice that $\theta$ is negative for $\varepsilon < 1/2$)
with the product $a = \theta \sqrt{2c}$ finite. 
Introducing these assumptions into (\ref{Eqtheta1}), 
we determine $\varepsilon$ as a function of $a$: 

\begin{equation}
\label{afonc}
\varepsilon = \frac{e^{a} \left( 1-a \right) - 
1}{2 \, (\cosh a - a \sinh a - 1)}.
\end{equation}

\noindent The relation between $\alpha, \theta$ 
and $a$ follows from (\ref{Eqc1}) and (\ref{afonc}): 

\begin{equation}
\label{thetafonc}
\frac{1}{\alpha} = \frac{\exp \left( - 
\theta^2/2 \right)}{\theta^3 \sqrt{2\pi}} \left\{  \frac{a^2 \, 
\left(\sinh a - a \right)}{\cosh a - a \sinh a - 1} \right\}. 
\end{equation}

\noindent $a$ and $\theta$ are increasing functions of 
$\varepsilon$ as expected.
For a symmetric distribution of targets ($\varepsilon = 1/2$)
then $a = 0$ corresponding to a vanishing threshold.
Conversely, if all the targets are $+1$ ($\varepsilon = 0$),
the threshold diverges to $- \infty$.
For finite $\varepsilon < 1/2$, the absolute value of 
the threshold is an increasing function of $\alpha$. From  
equation (\ref{thetafonc}) we obtain the development
$\theta^2 = 2 \ln \alpha + O(\ln \ln \alpha)$.
Notice that neglecting $\ln \ln \alpha$
with respect to $\ln \alpha$ is an approximation 
only valid for large enough $\alpha$ ($\alpha > 
10^{10}$). As was already pointed out in~\cite{WeSa}, 
this behaviour cannot be deduced by solving 
the equations (\ref{Eqc1}) and (\ref{Eqtheta1}) 
numerically. 

The training error ${\cal E}_t (\alpha, 
\varepsilon)$~(\ref{Et1}) with $\kappa = 0$ in 
the limit $\alpha \rightarrow + \infty$ is then:

\begin{equation}
\label{Etfonc}
{\cal E}_t (\alpha,\varepsilon) \simeq \varepsilon - 
\frac{\exp \left( - \theta^2/2 \right)}{\theta \sqrt{2\pi}} 
\left\{  \frac{\sinh a - a}{\cosh a - a \sinh a - 1} \right\}.
\end{equation}

\noindent Using equations~(\ref{thetafonc}) and 
(\ref{Etfonc}), we deduce:
  
\begin{equation}
\label{EqEt}
{\cal E}_t (\alpha,\varepsilon) \simeq \varepsilon - 
\frac{\theta^2}{\alpha \, a^2(\varepsilon)} \simeq 
\varepsilon - \frac{\ln \alpha}{\alpha} \, 
\frac{2}{a^2(\varepsilon)}
\end{equation}

\noindent where $a(\varepsilon)$ is the inverse function of 
$\varepsilon(a)$ given by (\ref{afonc}). 

Consider now the Gardner potential with finite $\kappa$.
Although, a solution of equations (\ref{Eqc1}) 
and (\ref{Eqtheta1}) under the assumption that $c 
\rightarrow 0$ with finite $\theta$ in the limit 
$\alpha \rightarrow + \infty$ exists, it does {\it not} 
correspond to the correct extremum of $g$~(\ref{foncg}).  
It is however worth to examine it.
The corresponding value of $\theta$ as a function 
of $\varepsilon$ and $\kappa$ follows form (\ref{Eqtheta1}), 
and the relation between $\alpha, \theta, \kappa$ and 
$c$ from (\ref{Eqc1}). We find:

\begin{eqnarray}
\varepsilon & = & \frac{1}{1 + \exp (2 \kappa \theta)},\\
\frac{1}{\alpha} & = & \frac{2 \varepsilon \, (2c)^{3/2}}{3 
\sqrt{2\pi}} \exp \left( - \frac{(\kappa + \theta)^2}{2} \right). 
\end{eqnarray}

As $\sqrt{2c} < \kappa$, the training error given by 
equation (\ref{Et1}) writes:

\begin{equation}
\label{WrongEt}
{\cal E}_t (\alpha,\varepsilon) =
\varepsilon + (1-2\varepsilon) \int_{-\theta}^{+ \infty} Dy
\end{equation}

\noindent and is larger than $\varepsilon$ for any 
finite $\theta$. Notice that this (incorrect) solution does not 
satisfy the condition (\ref{condition}) necessary 
for the TLA to converge.  

In fact, the correct training error corresponds  
to a solution with finite gap ($\sqrt{2c} 
\rightarrow 2 \kappa$) and a diverging 
threshold ($\theta \rightarrow - \infty$)
in the large $\alpha$ limit. 
Defining $\delta \equiv 2 \kappa - \sqrt{2c}$, 
and keeping only the leading terms, 
equations~(\ref{Eqc1}), (\ref{Eqtheta1}) and 
(\ref{Et1}) for $\kappa > 0$ give:

\begin{eqnarray}
\frac{\varepsilon}{1 - \varepsilon} & \simeq &  - 
\frac{\exp \left( \delta (\theta + \kappa) 
\right)}{2 \kappa (\theta + \kappa)}, \\ 
\frac{1}{\alpha} & \simeq & \frac{2 \kappa (1 - 
\varepsilon)}{(\theta+\kappa)^2 \sqrt{2 \pi}} \exp 
\left( - \frac{(\theta+\kappa)^2}{2} \right), \\
\label{EqEtkappa}
{\cal E}_t (\alpha,\varepsilon) & \simeq & \varepsilon 
- \frac{1}{(2\kappa)^2} \, \frac{1}{\alpha}.
\end{eqnarray}

\noindent The neglected terms are of the order  
$O(\exp(-2 \kappa \sqrt{2 \ln \alpha} + \ln \ln \alpha))$,
which are only negligible if $\kappa$ is finite. 
The prefactor $1/(2 \kappa)^2$ in~(\ref{EqEtkappa}),  
that diverges when $\kappa \rightarrow 0$, reflects 
the existence of the different behaviours for vanishing 
and for finite $\kappa$. 
 
This second solution only exists for 
bounded potentials. The Perceptron 
and the AdaTron potentials diverge for 
$\lambda \rightarrow -\infty$, and  
the corresponding training errors become 
larger than $\varepsilon$ in the large 
$\alpha$ limit. Thus, if these learning algorithms 
were used to train the hidden perceptrons, 
the TLA would not converge. 

Although the case of unbiased targets (i.e. 
$\varepsilon = 1/2$) is not essential for 
our study, we include here the corresponding analytic results 
for the sake of completeness. In this case, 
the free energy $g$~(\ref{foncg}) is invariant with 
respect to the threshold symmetry $\theta 
\leftrightarrow -\theta$. Thus, $\theta = 0$ is a trivial 
extremum of $g$. However, as already discussed by 
West and Saad in~\cite{WeSa}, two new solutions breaking 
the threshold symmetry appear above a given training set 
size $\alpha_{\theta}$. The analytical expression of 
$\alpha_{\theta}$ may be deduced under the assumption 
that the two different solutions appear continuously 
at $\alpha_{\theta}$, as in usual second order phase transitions, 
through a series expansion of the free energy in powers 
of $\theta$: 

\begin{equation}
g(\alpha,\varepsilon,c,\theta) = g(\alpha,\varepsilon,c,0) 
+ \frac{\theta^2}{2} \left. \frac{\partial^2 g}{\partial 
\theta^2} \right|_{\theta = 0} + \frac{\theta^4}{24} \left. 
\frac{\partial^4 g}{\partial \theta^4} \right|_{\theta = 0}.
\end{equation}

\noindent Due to the symmetry, the odd derivatives with 
respect to $\theta$ vanish. 
The condition:

\begin{equation}
\frac{\partial^{\, 2} g}{\partial \theta^{\, 2}} = 0 = 
\int_{\kappa-\sqrt{2c}}^{\kappa} y \, (\kappa - y) Dy
\end{equation}

\noindent defines $\sqrt{2c}$ at the transition. 
The size $\alpha_{\theta}$ satisfies:

\begin{equation}
\alpha_{\theta} = \left( \int_{\kappa-\sqrt{2c}}^{\kappa} 
(\kappa - y)^2 Dy \right)^{-1}
\end{equation}

\noindent and the two new solutions that appear for 
$\alpha > \alpha_{\theta}$ correspond to a threshold 
$\theta_{\pm} \sim \pm \sqrt{\alpha - \alpha_{\theta}}$. 
Notice that the usual stability criterion for second order 
phase transitions, ${\partial^{\, 4} g}/
{\partial \theta^{\, 4}} > 0$, cannot be directly applied 
here because we have two order parameters. Taking into account the 
leading corrections to $c$, proportional to $\theta^2$, 
it is straightforward to verify that the solutions with 
finite threshold are stable. 

\subsection{Kuhn-Tucker cavity method}

In order to circumvent the RS approximation, 
we determine the training error ${\cal E}_t (\alpha, 
\varepsilon)$ using the Kuhn-Tucker (KT) cavity method 
proposed by Gerl and Krey~\cite{GeKr}, that we generalize 
here to the case of a perceptron with a threshold 
learning a training set with a biased probability of 
targets given by~(\ref{prob}). 
Contrary to the RS solution, this cavity method has 
been shown to overestimate the training error~\cite{GeKr}. 
Consequently, the results allow us to deduce an upper 
bound for the number of perceptrons needed by the 
tilinglike procedure to converge.

The KT cavity method allows to determine the properties
of the perceptron by analyzing self-consistently its
response to the introduction of a new pattern into
the training set.
It is particularly adapted to study the properties 
of the Gardner potential~(\ref{potential}) because  
it is based on the fact that the weights minimizing the 
corresponding cost function are a (conveniently normalized) 
linear combination of the patterns with stability 
$\kappa$, which are called {\it support vectors}.
 
Let us assume that the perceptron has learned the training 
set and that the value of the cost function is $E$. 
This is the number of examples with stability 
smaller than the margin $\kappa$. 
The support vectors belong to the subset of 
$\alpha N - E$ remaining examples that do not 
contribute to the cost. The perceptron's weights 
may be expressed as follows: 
 
\begin{equation}
\label{eq_J}
{\bf J} = \frac{1}{N} \sum_{\mu \in \{\alpha N - E \} }
\tau^{\mu} a^{\mu} \, {\bf x}^{\mu}
\end{equation}
 
\noindent with $a^{\mu} > 0$ for $\lambda^{\mu} = 
\kappa$, and $a^{\mu} = 0$ for $\lambda^{\mu} > 
\kappa$. These are the so-called Kuhn-Tucker conditions. 
Defining $a^{\mu} = 0$ for examples with $\lambda^{\mu} < 
\kappa$, the normalization of the weights imposes:
 
\begin{equation}
\label{norm}
1 = {\bf J} \cdot {\bf J} = \frac{1}{N}
\sum_{\mu=1}^{\alpha N} \tau^{\mu} a^{\mu} \, {\bf x}^{\mu}
\cdot {\bf J} = \frac{1}{N} \sum_{\mu = 1}^{\alpha N} a^{\mu}
\left( \kappa + \tau^{\mu} \theta \right).
\end{equation}
 
As usual with cavity methods, we introduce a new 
example ${\bf x}^0$ with target $\tau^0$, drawn respectively
with the same probability densities as the other 
inputs and targets in the training set.
Before any modification, as the pattern $0$ is 
uncorrelated with the direction ${\bf J}$ and its 
components are assumed to have a gaussian distribution, its 
projection onto ${\bf J}$ has a gaussian probability. 
Therefore, the joint probability distribution of the target 
$\tau^0$ and the stability $\tilde{\lambda}^0= 
\tau^0 ({\bf x}^0 \cdot {\bf J} - \theta)$ 
before learning is: 

\begin{equation}
\Pi \left( \tilde{\lambda}^0,\tau^0 \right) =
\frac{P(\tau^0)}{\sqrt{2 \pi}} \exp \left( -
\frac{(\tilde{\lambda}^0 + \tau^0 \theta)^2}{2} \right)
\end{equation}
 
\noindent where $P(\tau^0)$ is defined by~(\ref{prob}). 
We assume a single ground state and we calculate the 
necessary adjustments of the weights ${\bf J}$ in order 
to obtain self-consistent equations for the cost function 
as a function of $\alpha$. 

If $\tilde{\lambda}^0 \geq \kappa$, no learning 
is needed, as the new example does not 
contribute to the cost.
If $\tilde{\lambda}^0 < \kappa$, 
two different situations may occur. 
Either the distance of the new example to the 
hyperplane is too large and the perceptron is 
unable to learn it, or the example is close 
enough and can be learned. The natural strategy to 
minimize the cost function is 
to include the new example in the 
subset of support vectors only if $\kappa - \sqrt{2c} < 
\tilde{\lambda}^0 < \kappa$, where  
$\sqrt{2c}$ is a positive quantity which 
has to be determined self-consistently. 
Otherwise, the weights are not modified and the 
new example is left in the 
subset of examples contributing to the cost.
We are left with the problem of determining the 
perturbation on the weights such that examples 
with $\kappa - \sqrt{2c} < \tilde{\lambda}^0 
< \kappa$ become support vectors after learning. 
As a first step, this can be obtained by
taking  $a^0 = \kappa - \tilde{\lambda}^0$.
However, this modifies the stabilities of the other 
support vectors. The coefficients $a^{\mu} > 0$ 
($\mu \geq 1$) must be corrected by a small amount 
to compensate for this perturbation.
This correction in turn modifies the stability of the 
new example $0$, and $a^0$ has to be corrected. 
After a full summation of the contributions, 
Gerl and Krey~\cite{GeKr} have shown that the 
correct value of $a^0$ is:

\begin{equation} 
\label{eq_a0}
a^0 = \frac{\kappa - \tilde{\lambda}^0}{1-\alpha P(a^{\mu} > 0)}
\end{equation}

\noindent where $P(a^{\mu} > 0)$ is the probability 
that $a^{\mu} > 0$. This probability is determined 
assuming that the new example is equivalent to the others: 
 
\begin{equation}
\label{eq_Pamu}
P ( a^{\mu} > 0) \equiv P(a^0 > 0) = \sum_{\tau^0 = \pm 1} 
\int_{\kappa - \sqrt{2c}}^{\kappa} \Pi \left( 
\tilde{\lambda}^0,\tau^0 \right) d\tilde{\lambda}^0.
\end{equation}
 
Having specified the learning procedure, we are
able to determine $\sqrt{2c}$ and $E$ self-consistently. 
First of all, the normalization of the weights given 
by equation~(\ref{norm}), may be written as follows:

\begin{equation}
\label{eqnorm}
1 = \alpha \sum_{\tau^0 = \pm 1} \int_{-\infty}^{+\infty}
a^0 \left(\kappa + \tau^0 \theta \right) \Pi \left(
\tilde{\lambda}^0,\tau^0 \right) d\tilde{\lambda}^0
\end{equation}

\noindent with $a^0$ given by (\ref{eq_a0}) 
for $\kappa-\sqrt{2c} < \tilde{\lambda}^0 
< \kappa$ and $a^0=0$ elsewhere. Combining 
equations~(\ref{eq_Pamu}) and (\ref{eqnorm}), 
we obtain:

\begin{eqnarray}
\label{eq2c}
1 & = & \alpha \,  (1 - \varepsilon) \int_{\kappa - 
\sqrt{2c}+\theta}^{\kappa+\theta} \left( 1 + (\kappa + 
\theta)(\kappa + \theta - y) \right) Dy\\
\nonumber & &  + \, \alpha \, \varepsilon \int_{\kappa
- \sqrt{2c}-\theta}^{\kappa-\theta} \left( 1 + (\kappa 
- \theta)(\kappa - \theta - y) \right) Dy.
\end{eqnarray}
 
\noindent This equation, which determines $\sqrt{2c}$ 
for a fixed threshold $\theta$, is slightly different 
from the RS result~(\ref{Eqc1}). The cost function $E$ 
is determined assuming that it remains unchanged (to 
order $\sqrt{N}$) upon learning the new example. Thus, 
the cost per example writes:
 
\begin{eqnarray}
\nonumber \frac{E}{\alpha N} & = & \sum_{\tau^0 = \pm 1} \int_{-\infty}^{
\kappa - \sqrt{2c}} \Pi \left( \tilde{\lambda}^0,\tau^0
\right) d\tilde{\lambda}^0\\
\label{eqet}
& = & (1-\varepsilon) \int_{-\infty}^{\kappa-\sqrt{2c}+\theta}
Dy + \varepsilon \int_{-\infty}^{\kappa-\sqrt{2c}-\theta} Dy.
\end{eqnarray}
 
Notice that when $\kappa - \sqrt{2c} < 0$,
$E/(\alpha N)$~(\ref{eqet}) represents the fraction of
training errors ${\cal E}_t (\alpha, \varepsilon)$
and is similar to~(\ref{Et1}).
The threshold $\theta$ may be optimized 
in order to minimize the cost function:
 
\begin{equation}
\label{eq_theta}
\frac{\partial {E}}{\partial \theta} = 0.
\end{equation}

In the following, we solve~(\ref{eq2c}) 
and~(\ref{eq_theta}) in the large $\alpha$ limit. 
First of all, we consider the case $\kappa = 0$. 
In this case, $E/(\alpha N)$ (equation~(\ref{eqet})) 
is the training error ${\cal E}_t$. As for the RS 
calculation, we may assume 
$\sqrt{2c} \ll |\theta|$ and $a = \theta \sqrt{2c}$ 
finite. We obtain the following equations:

\begin{eqnarray}
\label{eqaKT}
\frac{\varepsilon}{1 - \varepsilon} & \simeq \hskip 0.3cm & 
e^{2a} \left( \sqrt{4a^2+1} - 2a \right),\\
\hskip 0.5cm \frac{1}{\alpha} & \simeq & a \left\{ 
(1 - \varepsilon) \, e^a + \varepsilon \, e^{-a} \right\} 
\frac{\exp \left(- \theta^2/2 \right)}{ \theta \sqrt{2 \pi}},\\ 
\label{eqEtKT}
{\cal E}_t (\alpha, \varepsilon) & \simeq & \varepsilon 
- \frac{F(a)}{\alpha} \simeq \varepsilon - \frac{1}{\alpha} 
\, \frac{\left( 1 + 2a - \sqrt{4a^2+1} \right)}{a  
\left( 1 + \sqrt{4a^2+1} - 2a \right)}. 
\end{eqnarray}

These results differ from those obtained with 
the RS calculation (Equations (\ref{afonc}), 
(\ref{thetafonc}) and (\ref{EqEt})). 

In the case of finite margin $\kappa$,  
the pertinent assumptions in the large 
$\alpha$ limit are $\sqrt{2c} \rightarrow 
2 \kappa$ with $\delta = 2 \kappa - \sqrt{2c}$ 
and $\theta \rightarrow - \infty$. With these, 
here again $E/(\alpha N)$~(\ref{eqet}) is the training 
error, and we get: 

\begin{eqnarray}
\frac{\varepsilon}{1 - \varepsilon} & \simeq \hskip 0.3cm & 
\frac{4 \exp(\delta (\theta+\kappa))}{(\theta + \kappa)^2},\\
\hskip 0.5cm \frac{1}{\alpha} & \simeq & \frac{8 \kappa 
(1 - \varepsilon)}{(\theta + \kappa)^2 \sqrt{2\pi}} \exp 
\left( - \frac{(\theta + \kappa)^2}{2} \right),\\
\label{eqEtKTkappa}
{\cal E}_t (\alpha, \varepsilon) & \simeq & \varepsilon + 
\frac{1}{2 \kappa \alpha (\theta + \kappa)} \simeq 
\varepsilon - \frac{1}{2 \kappa \alpha \sqrt{2 \ln \alpha}}. 
\end{eqnarray}

It is worth to point out that even within the KT 
cavity method, the training error satisfies 
the convergence conditions~(\ref{condition}) 
and~(\ref{stop}).  

The main conclusion of this section is that the 
TLA converges provided that the hidden perceptrons 
are trained through the minimization of a cost 
function with a bounded potential. 
The Gardner potential~(\ref{potential}) satisfies 
this constraint. The asymptotic behaviours 
of the training error in the large $\alpha$ limit, 
calculated for $\kappa = 0$ and $\kappa \neq 0$ 
using two different approaches are used in the 
following sections to characterize the storage capacity
of the constructive algorithm.

\section{Number of hidden perceptrons in the large 
$\alpha$ limit}
\label{sec:k(alpha)}

We assume that the probability distribution of the 
targets $\tau^{\mu}$ in the training set is symmetric, 
given by (\ref{prob}) with $\varepsilon= 1/2$, so 
that the training error of the first perceptron is 
$\varepsilon_t^1 = {\cal E}_t (\alpha, 1/2)$. 
Considering iteratively the relationship between the
training errors of two consecutive
perceptrons~(\ref{relationEt}) yields:

\begin{equation}
\circ_{k} f_{\alpha} (1/2) = \underbrace{f_{\alpha}
 \circ \cdots \circ f_{\alpha}}_{k \
{\rm times}} \ (1/2) = 0
\end{equation}

\noindent where $f_{\alpha} (\varepsilon)$ stands for 
${\cal E}_t (\alpha,\varepsilon)$, the symbol $\circ$ 
for the composition of functions and $k$ is the number 
of perceptrons necessary for convergence of the 
TLA algorithm. 

\begin{figure}
\centerline{\psfig{figure=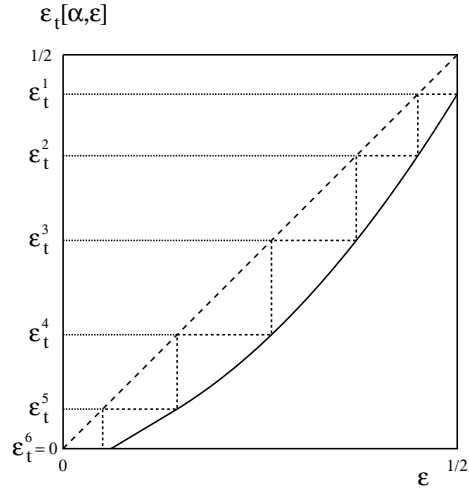,height=7cm}}
\caption{
\label{epsit}
Evolution of the successive training errors. 
The full curve corresponds to the training error 
${\cal E}_t (\alpha,\varepsilon)$ of a  
perceptron with biased targets.
The first training error $\varepsilon_t^1$ is 
given by ${\cal E}_t (\alpha,1/2)$ and the 
following ones by the relation $\varepsilon_t^{i+1} 
= {\cal E}_t (\alpha,\varepsilon_t^i)$.
In this case, the learning algorithm converges with 
six perceptrons. }
\end{figure}

The evolution of the training errors of the successive 
perceptrons is schematically represented on figure 
\ref{epsit} for an arbitrary function ${\cal E}_t 
(\alpha,\varepsilon)$, where the tilinglike algorithm 
is shown to converge in six steps, i.e. $k = 6$. 

We are interested in the limit of large training set 
sizes ($\alpha \rightarrow + \infty$). In this limit, 
the training error ${\cal E}_t (\alpha,\varepsilon)$ 
is close to $\varepsilon$:

\begin{equation}
\label{fonch}
{\cal E}_t (\alpha,\varepsilon) \simeq 
\varepsilon - h(\alpha,\varepsilon)
\end{equation}

\noindent with $h(\alpha,\varepsilon)$ a function that 
vanishes in the limit $\alpha \rightarrow +\infty$.
Notice that those cost functions that do not satisfy 
condition (\ref{condition}) for all $\alpha$ are useless  
in this limit, since the error reduction at each step 
$\varepsilon_t^{i+1} - \varepsilon_t^i = - 
h(\alpha,\varepsilon_t^i)$ vanishes at some finite 
value of $\alpha$. For larger values of $\alpha$ 
it becomes positive, and the TLA does not converge. 
In the preceding section we showed that 
the Gardner potential both with vanishing 
and finite margin $\kappa$ has 
$h(\alpha,\varepsilon) > 0$ (see equations 
(\ref{EqEt}) and (\ref{EqEtkappa})) 
and satisfies condition (\ref{condition}).

As $h(\alpha,\varepsilon)$ vanishes in the limit 
$\alpha \rightarrow + \infty$, we can 
guess that the number $k(\alpha)$ diverges.  
In this limit we can introduce the continuum approximation, 
replacing $i/k$ by the real-valued variable $x$. 
Then, the error reduction at each step is given by: 

\begin{equation}
\label{derivee}
\varepsilon_t^{i+1} - \varepsilon_t^i \simeq \frac{1}{k} 
\frac{d\varepsilon}{dx} =  - h(\alpha, \varepsilon).  
\end{equation}

\noindent After integration of both sides of the equation 
$d\varepsilon/h(\varepsilon,\alpha) = - k \, dx$ 
at constant $\alpha$, from $\varepsilon = 1/2$ and 
$x = 0$ to $\varepsilon = 0$ and $x = 1$, we obtain:

\begin{equation}
\label{kalpha}
k (\alpha) \simeq  \int_{0}^{1/2} \frac{d \varepsilon}{ 
h (\alpha,\varepsilon)} = \int_{0}^{1/2} \frac{d 
\varepsilon}{\varepsilon - {\cal E}_t (\alpha,\varepsilon)}.
\end{equation}

\noindent Equation~(\ref{kalpha}) gives the asymptotic 
behaviour of the number of hidden perceptrons necessary 
for the tilinglike algorithm to converge in the limit 
$\alpha \rightarrow +\infty$. It depends 
on the cost function used to train the 
perceptrons through ${\cal E}_t (\alpha,\varepsilon)$.
The storage capacity $\alpha_c(k)$ of the TLA is 
then obtained through the inversion of $k(\alpha)$.

Hereafter we consider the case where the hidden 
perceptrons are trained with the Gardner cost function, 
using the results of the preceding section. 

We determine first the number of hidden units 
obtained when the perceptrons minimize the 
number of training errors, that is, 
the Gardner cost function with $\kappa = 0$. 
Inserting into~(\ref{kalpha}), the result~(\ref{EqEt}) 
obtained within the RS approximation, we obtain:
 
\begin{equation}
k^{RS} (\alpha) \simeq \int_{0}^{1/2} \frac{d\varepsilon}{
\varepsilon - {\cal E}_t (\alpha,\varepsilon)}
\simeq \frac{\alpha}{2 \, \ln \alpha} \int_{0}^{1/2}
a^2(\varepsilon) \,  d \varepsilon \simeq 0.475 \,
\frac{\alpha}{\ln \alpha}
\end{equation}
 
\noindent where $a(\varepsilon)$ is given by~(\ref{afonc}). 
From this result, we deduce the storage capacity 
in the limit of a large number of hidden perceptrons:
 
\begin{equation}
\label{alphaRS}
\alpha_c^{RS} (k) \simeq 2.11 \, k \ln k. 
\end{equation}

\noindent Surprisingly, the capacity of the TLA 
scales with $k$ like the upper bound for the 
parity machine with the same number of hidden 
units, and only the prefactor is overestimated.

Using the result (\ref{eqEtKT}) obtained with the 
KT cavity method, that overestimates the perceptron's 
training error, we get:

\begin{equation}
k^{KT}(\alpha) \simeq \alpha \int_{0}^{1/2}
\frac{d\varepsilon}{F(a(\varepsilon))} \simeq 1.082 \, \alpha
\end{equation}
 
\noindent where $F(a)$ is defined in (\ref{eqEtKT}) and 
$a(\varepsilon)$ is given by (\ref{eqaKT}). 
The corresponding storage capacity is: 

\begin{equation}
\label{alphaKT}
\alpha_c^{KT} (k) \simeq 0.924 \, k.
\end{equation}

We find that $\alpha_c^{KT} < \alpha_c^{RS}$ as expected.
The behaviour of the storage capacity, obtained with
the Kuhn-Tucker cavity method is linear in $k$.
This suggests that including replica symmetry breaking
in the replica calculation may modify the $k \ln k$ behaviour
to one proportional to $k (\ln k)^{\nu}$ with 
$0 \leq \nu \leq 1$. However, as the actual training 
error of the perceptrons seems closer to the RS  
solution than to the Kuhn-Tucker cavity result~\cite{GeKr}, 
we expect $\nu$ to be close to $1$.

In the following we consider the parity machine obtained  
when the perceptrons are trained using the Gardner 
cost function with a finite margin $\kappa$. We 
get: 

\begin{equation}
\label{kRSKT}
k^{RS}(\alpha,\kappa) \simeq 2 \kappa^2 \alpha, 
\hskip 0.5cm {\rm and} \hskip 0.5cm  k^{KT}
(\alpha,\kappa) \simeq \kappa \alpha \sqrt{2 \ln \alpha}.
\end{equation}

\noindent After inversion of (\ref{kRSKT}), the 
capacities deduced within the two approximations are:

\begin{equation}
\label{alphaRSKT}
\alpha_c^{RS} (k,\kappa) \simeq \frac{k}{2\kappa^2}, 
\hskip 0.5cm {\rm and} \hskip 0.5cm \alpha_c^{KT} 
(k,\kappa) \simeq \frac{k}{\kappa \sqrt{2 \ln k}} 
\end{equation}

\noindent respectively. Here again, the behaviours 
of $k(\alpha)$ and $\alpha_c (k)$ obtained with the 
RS approximation and with the Kuhn-Tucker 
cavity method differ. In both cases, the value of 
$\kappa$ only affects the prefactor but not the 
scaling with $\alpha$ or $k$. Consistently, the 
prefactor of $\alpha_c$ diverges for $\kappa 
\rightarrow 0$, where the expressions 
(\ref{kRSKT}) and (\ref{alphaRSKT}) have to be 
replaced by (\ref{alphaRS}) and (\ref{alphaKT}) 
respectively, as the functional dependence of the 
storage capacity with $k$ is different for 
$\kappa = 0$.

Imposing a finite margin dramatically decreases 
the capacity of the TLA. More precisely, the 
exponents $\nu$ of the logarithmic factor differ, 
depending on the approximations (RS and KT cavity 
method), in both $\kappa$-regimes ($\nu^{RS}(\kappa 
= 0) = 1$, $\nu^{RS}(\kappa > 0) = 0$, 
$\nu^{KT}(\kappa = 0) = 0$ and $\nu^{KT}(\kappa
> 0)= - 1/2$).

It is interesting to compare the exponents  
determined analytically within the RS approximation, 
to those obtained by West and 
Saad~\cite{WeSa2} through a numerical 
iteration over the successive perceptrons'  
training errors. 
For $\kappa = 0$, they obtain $\nu$ close to $1$ 
($n_e = 1.070$ and $1.049$, and $n_l = 1.079$ 
and $1.062$, for $k = 1000$ and $4000$ respectively 
(table 3 in~\cite{WeSa2})) in very good agreement 
with our result $\nu^{RS}(\kappa = 0) = 1$. 
In the case of finite $\kappa$, West and Saad 
find that the exponent decreases with increasing $\kappa$ 
(figure 13 left in~\cite{WeSa2}). Our result 
(\ref{alphaRSKT}) shows that the exponent does 
not depend on $\kappa$, only the prefactor does. 
The dependence found numerically is probably due 
to higher order corrections, that behave like 
$O(\exp ( - 2 \kappa \sqrt{2 \ln \alpha} + 
\ln \ln \alpha ))$. These terms, which are less and less 
negligible when approaching $\kappa = 0$, 
hinder the determination of the power-law exponent 
in the asymptotic regime $\alpha \rightarrow +\infty$. 
Remarkably, the RS and KT exponents $\nu^{RS}$ and 
$\nu^{KT}$ provide correct upper and lower bounds 
for the exponent obtained numerically within the one-step replica 
symmetry breaking approximation (figure 13 right in~\cite{WeSa2}).

\section{Conclusion}
\label{sec:concl}

We determined {\it analytically} the typical number of hidden 
units needed by a simple constructive procedure, the Tilinglike 
Learning Algorithm proposed in~\cite{BiOp}, to build a parity
machine. The number of hidden units depends strongly on 
the asymptotic properties of the learning algorithm used 
to train them. 

We showed that the cost function minimized by the 
hidden perceptrons has to be bounded. This rules 
out, in particular, the Perceptron or the AdaTron 
learning algorithms, as with these 
the training error cannot decrease beyond a finite 
value that depends on the training set size and on 
the bias of the target's distribution. This is so 
because the hidden perceptrons have to learn 
highly biased output distributions. In the asymptotic 
regime, large thresholds are needed to minimize the 
training error as, loosely speaking, such solutions 
allow to classify correctly most patterns of the 
majority class. In such solutions, 
a non-negligible fraction of patterns have large 
negative stabilities. If the cost function is 
unbounded for $\lambda \rightarrow -\infty$, 
it favours solutions with small thresholds, which 
have large training errors. 
With bounded potentials, like the counting functions 
used in the Gardner cost function, solutions with 
large thresholds exist. 

We deduced the properties of a perceptron with threshold, 
learning targets drawn with a biased distribution, 
trained with the Gardner cost function with and without 
margin. In particular, solutions such that the training 
error is smaller than the bias always exist. This is a 
condition necessary for the TLA to converge.
The asymptotic behaviour of the learning curves ${\cal E}_t 
(\alpha,\varepsilon)$ was determined through a replica calculation 
assuming replica symmetry, and also using the Kuhn-Tucker 
cavity method. The former approximation underestimates the 
training error, while the latter overestimates it. The main 
results are the expressions~(\ref{EqEt}), (\ref{EqEtkappa}), 
(\ref{eqEtKT}) and (\ref{eqEtKTkappa}) relating the training 
error of the perceptron ${\cal E}_t (\alpha,\varepsilon)$ to 
the bias $\varepsilon$ of the target distribution. Closer 
inspection of equations (\ref{EqEt}) and (\ref{EqEtkappa}) 
shows that the error reduction ${\cal E}_t (\alpha,\varepsilon) 
- \varepsilon$  at large $\alpha$ is larger if $\kappa=0$ than 
for $\kappa >0$.   

These results allow us to find analytically the number of 
units $k(\alpha)$ needed by the constructive procedure to 
converge in the large $\alpha$ limit. As expected, the 
smallest $k(\alpha)$ is obtained when the hidden perceptrons 
minimize their training errors, which corresponds to the 
Gardner cost function with $\kappa = 0$. Nevertheless, it is 
worth to study also the case with $\kappa > 0$, which is 
interesting in noisy applications. The storage capacity 
$\alpha_c(k)$ of the TLA is obtained through the inversion 
of $k(\alpha)$. Our results have been obtained under the 
simplifying assumption that the targets the successive 
perceptrons have to learn are uncorrelated. This hypothesis 
has been shown to be a good approximation~\cite{WeSa2} in 
the limit of large training sets considered here.

In the limit of large $k$ we find $\alpha_c^{RS} (k) 
\simeq 2.11 \, k \ln k$ within the RS approximation. It 
is interesting to compare this algorithm-dependent storage 
capacity to the storage capacity of a parity machine with 
the same number of hidden perceptrons. The latter is 
independent of the learning algorithm. Geometric 
arguments~\cite{MiDu} and a replica calculation where the 
permutation symmetry among hidden units has to be 
broken~\cite{XKO}, both lead to $\alpha_c = k \ln k / \ln 2$. 
It is surprising that, although we disregarded the 
correlations between perceptrons and assumed replica-symmetry, 
which both lead to an overestimation of the storage capacity,
 we find the same leading behaviour. Only the prefactor is 
overestimated. In fact, the permutation symmetry only arises 
when the perceptrons are trained simultaneously. As it is 
absent in the case of the incremental construction, the 
consequence of the RS approximation is less dramatic 
than in~\cite{XKO}.  

As the Kuhn-Tucker cavity method provides an upper bound 
to the perceptron's training error, it allows to determine 
a lower bound for the TLA storage capacity. This bound scales 
linearly with the number of hidden units, suggesting that 
a calculation including full replica symmetry-breaking may 
change the power-law of the logarithmic factor. We expect 
that $\alpha_c \sim k (\ln k)^{\nu}$ with $0 \leq \nu \leq 1$. 

\section*{Acknowledgements}

It is a pleasure to thank K. Y. Michael Wong for 
clarifying comments about the KT cavity method.


\end{document}